# Single-pixel imaging based on deep learning

Kai Song[1], Yaoxing Bian[1*], Ku Wu[2], Hongrui Liu[1], Shuangping Han[1], Jiaming Li[1], Jiazhao Tian[1], Chengbin Qin[1,3], Jianyong Hu[1,3*], and Liantuan Xiao[1,3*]

1. College of Physics, College of Electronic Information and Optical Engineering, Taiyuan University of technology, Taiyuan 030024, China
2. University of Leeds, School of Computing, Leeds, LS2 9JT, UK
3. State Key Laboratory of Quantum Optics and Quantum Devices, Institute of Laser Spectroscopy, Shanxi University, Taiyuan 030006, China
* Correspondence: bianyaoxing@tyut.edu.cn, jyhu@sxu.edu.cn, xlt@sxu.edu.cn

**Abstract:** Single-pixel imaging can collect images at the wavelengths outside the reach of conventional focal plane array detectors. However, the limited image quality and lengthy computational times for iterative reconstruction still impede the practical application of single-pixel imaging. Recently, deep learning has been introduced into single-pixel imaging, which has attracted a lot of attention due to its exceptional reconstruction quality, fast reconstruction speed, and the potential to complete advanced sensing tasks without reconstructing images. Here, this advance is discussed and some opinions are offered. Firstly, based on the fundamental principles of single-pixel imaging and deep learning, the principles and algorithms of single-pixel imaging based on deep learning are described and analyzed. Subsequently, the implementation technologies of single-pixel imaging based on deep learning are reviewed. They are divided into super-resolution single-pixel imaging, single-pixel imaging through scattering media, photon-level single-pixel imaging, optical encryption based on single-pixel imaging, color single-pixel imaging, and image-free sensing according to diverse application fields. Finally, major challenges and corresponding feasible approaches are discussed, as well as more possible applications in the future.

## 1. Introduction

As a typical computational imaging technique, single-pixel imaging (SPI) utilizes a set of spatial patterns to illuminate the samples, then uses a bucket detector for data acquisition, and forms an image from the measurements by inversion algorithm[1-3]. Unlike traditional focal plane array detectors, SPI only adopts a high-sensitivity photodetector to collect echo signals, which makes it have significant advantages in detection sensitivity, broadening spectral response range, and reducing imaging costs[4]. In the past decade, SPI has successfully expanded its spectral response range from visible light[5] to ultraviolet[6], infrared[7] and even terahertz wavelengths[8, 9]. These advancements have significantly enhanced its application in medical imaging[10-12], biological imaging[13-15], and industrial manufacturing[16].

Imaging speed and quality are two pivotal benchmarks that SPI constantly pursues. However, obtaining an excellent image necessitates numerous patterns, which extend the time for both sampling and reconstruction. This poses significant challenges for SPI in large scenes or dynamic reconstruction. To solve this problem, numerous studies optimize patterns and algorithms to reduce imaging time[17-19]. Despite these efforts have partially reduced the imaging time, there are still practical limitations in the applications of SPI. For example, the universal patterns, such as random scattering basis[20], Hadamard basis[21], Fourier basis[22], and cosine transform basis[23], exhibit limited adaptability to specific tasks, necessitating the incorporation of more patterns to ensure optimal image reconstruction quality.

With the development of compressive sensing (CS) theory[24], SPI based on CS (CS-SPI) has attracted considerable attention for its elegant integration of optics, mathematics, and optimization theory[25]. It possesses the ability to perform SPI with fewer patterns, which is essential in improving imaging speed. However, to recover high-quality images under sub-Nyquist sampling, the CS-SPI algorithms are typically implemented within the framework of iterative convex optimization, demanding substantial computational resources[26-28]. The unpredictable processing time may range from seconds to minutes, and is highly dependent on the complexity of the scene.

Additionally, the degradation of image quality caused by sub-Nyquist sampling also limits its application.

Recently, to overcome the limitations of the aforementioned methods, deep learning[29] has been applied for patterns generation and image reconstruction in SPI[30-32]. In comparison to iterative methods, deep learning algorithms can achieve faster reconstruction speed and higher reconstruction quality. In the SPI based on deep learning (DL-SPI), deep learning algorithms can improve the low-quality images obtained by iterative algorithms[33], and can also directly transform measurements into images[34]. Moreover, these algorithms can perform advanced sensing tasks such as object classification, segmentation, and detection by directly analyzing measurements without image reconstruction. This approach is commonly known as image-free sensing. Currently, the researches on DL-SPI covers a wide range of domains, as shown in Figure 1, including super-resolution SPI, SPI through scattering media, photon-level SPI, optical encryption based on SPI, color SPI, and image-free sensing.

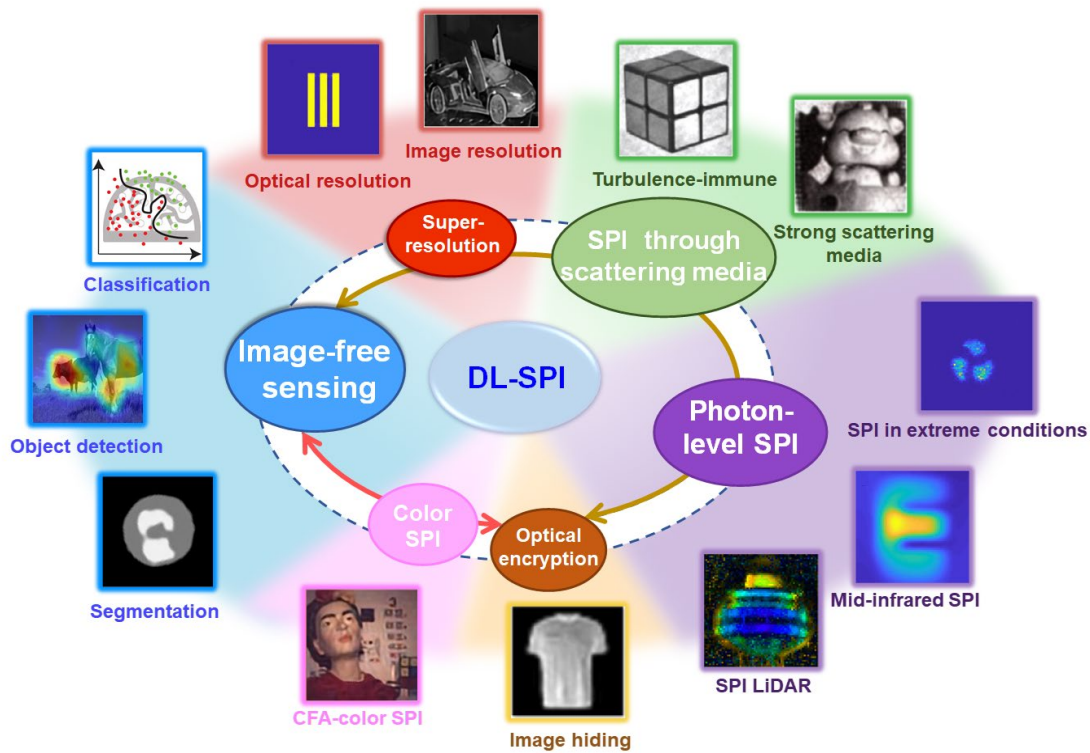

**Figure 1. Overview of DL-SPI.** The specific domains include super-resolution[35, 36], SPI through scattering media[37, 38], photon level SPI[7, 39, 40], optical encryption[41], color SPI[42], and image-free sensing[43-45].

This paper reviews the development of SPI with deep learning technology and look forward to the potential research directions of DL-SPI. The remainder of this paper is organized as follows: Section 2 describes the principles of two distinct SPI frameworks. Next, we initially present the fundamental principles of deep learning and subsequently elaborate on how deep learning algorithms are integrated with SPI. The following section classifies and summarizes the representative studies of DL-SPI from the perspective of application. Finally, we summarize this review and offer some perspectives on the potential future research avenues, advancement, and applications of DL-SPI. In addition, the existing challenges of DL-SPI also be discussed.

## 2. Single-pixel imaging

Scanning each pixel of the target image in turn by a single-pixel detector is a straightforward method for image capture[46, 47], but it only acquires information on one pixel at a time. It is an inefficient way of utilizing illumination light and typically requires a long scanning time. To overcome these limitations, SPI adopt spatially modulated illumination light or reflected light to perform a multi-modal scan of the object, and the echo signal is collected by a single-pixel detector (SPD). The distinctive scanning strategies of SPI can be further divided into structured detection and structured illumination[48]. This section provides a detailed introduction to the setup and principles of these two imaging strategies of SPI.

### 2.1 Structured detection single-pixel imaging

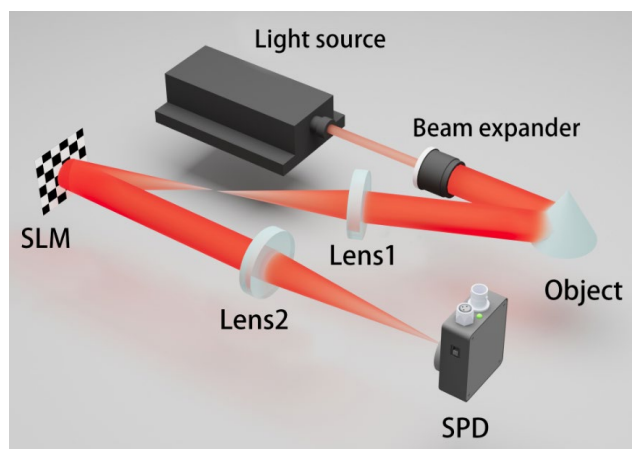

**Figure 2. Structured detection SPI.** An object is illuminated and imaged onto or through the spatial light modulator (SLM) pre-loaded with a set of patterns. The single-pixel detector (SPD) is used to collect the intensity for each pattern and the image is formed by inversion algorithm.

Structured detection SPI is initially proposed by Durate et al. based on CS, so it's also known as CS-SPI[49]. This approach adopts a post-modulation strategy, where objects are imaged onto or through the SLM pre-loaded with a set of patterns, such as digital micromirror device (DMD)[50] or liquid crystal spatial light modulator (SLM)[51]. Then the modulated light is collected by a SPD, as shown in Figure 2.

According to the principle of CS, a series of patterns displayed by SLM constitute the measurement matrix $\phi \in R^{M \times N}$, where $M$ represents the number of patterns and $N$ is the resolution of the target image. The measurements $y$ recorded by the SPD can be expressed as:

$$y_i = \phi_i x \ (i = 1,2,3 \ldots M) \quad (2\text{-}1)$$

where $x$ is the target image, represented by a one-dimensional (1D) column vector with a dimension of $N \times 1$. The dimension of the measurements $y$ is $M \times 1$. The sparsity of signals is widely recognized as a fundamental prerequisite for CS. However, the target images typically exhibit insufficient sparsity, necessitating transforming them into the sparse domain for a viable solution. If the target images can be sparsely represented by some transformation basis, $x = \psi s$, the measurements $y$ can be defined as:

$$y = \phi \psi s \quad (2\text{-}2)$$

where $\psi$ is the appropriate basis for projecting the original signal to the sparse domain. $s$ represents the sparse signal with most of the coefficients that are close or equal to zero.

For image reconstruction, the sparse signal $s$ can be obtained by performing the inverse operation on Eq. (2-2). The target image $x$ can be recovered through inverse domain transformation on $s$. However, due to the significantly larger number of unknowns in Eq. (2-2) compared to the number of equations (measurements), Eq. (2-2) has infinitely many solutions. It is difficult to recover the $N$-dimensional original signals directly from the $M$-dimensional measurements ($M \ll N$). Consequently, the researchers try to acquire the $x$ through a mathematical model with $L_0$ or $L_1$ norm constraints:

$$\begin{aligned}&\min_x ||x||_0 \textbf{ subject to } y = \phi\psi s \textbf{ or}\\&\min_x ||x||_1 \textbf{ subject to } y = \phi\psi s\end{aligned} \quad (2\text{-}3)$$

## 2.2 Structured illumination single-pixel imaging

In contrast to structured detection SPI, structured illumination SPI applies a SLM to generate illumination patterns with clear spatial distributions. The SPD collects the backscattered signal from the object illuminated by patterns (Figure 3). This imaging strategy is known as computational ghost imaging[52].

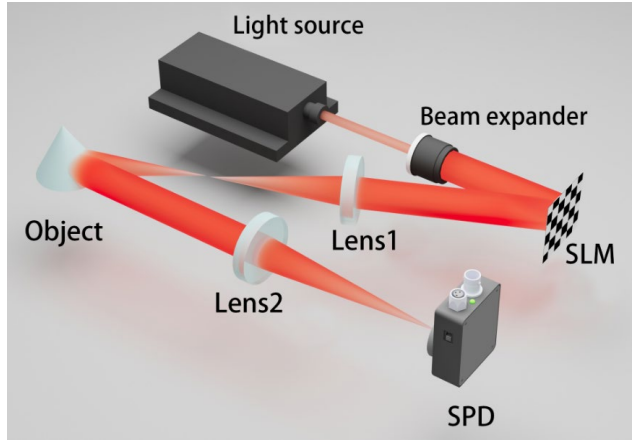

**Figure 3. Structured illumination SPI.** The SLM is used to project a sequence of illumination patterns onto an object and the SPD measures the backscattered signal.

Structured illumination imaging is proposed by Shapiro et al. in computational ghost imaging. In the study of Shapiro et al., the measurements $S$ is collected by a SPD can be expressed as:

$$S_m = \int I_{(x,y)}^m T_{(x,y)} dxdy \qquad (2\text{-}4)$$

where $I_{(x,y)}$ represents the patterns, $T_{(x,y)}$ is the target image, $(x,y)$ is the plane coordinate, and $m = 1,2,3 \ldots M$ is the number of measurements. According to the principle of computational ghost imaging, the image can be reconstructed by calculating the higher-order correlation between the patterns $I_{(x,y)}$ and the measurements $S$:

$$T(x,y)_{GI} = <\Delta S_m \Delta I_m> \qquad (2\text{-}5)$$

where $<\cdot>$ denotes mean value, $\Delta S_m = S_m - <\Delta S_m>$, $\Delta I_m = I_m - <I_m>$.

Comparing the two imaging strategies, it can be observed that structured detection SPI modulates the scattered light from the object, making it suitable for passive imaging. However, structured illumination utilizes illumination patterns to encode the object, which belongs to active imaging. Both of two imaging strategies can be employed not only for the recovery of 2D images but also for the reconstruction of 3D images. Time-of-flight, fourier transform profilometry, tomography and stereo vision are several main approaches conceived for 3D SPI. In the time-of-flight scheme, the continuous laser is replaced by a pulsed laser, and then the photon statistical histogram is plotted for each pattern. The photon count within the same time bin at different statistical histograms can be regarded as a set of measurements. The measurements associated with different time bins represent different depths. Finally, the reconstruction algorithm utilizes the measurements from different time bins to recover a series of 2D images, resulting in a 3D image cube[53]. For fourier transform profilometry, the 3D structure of the object can be reconstructed according to the deformed fringe pattern captured by the single pixel detector[54]. The single-pixel tomography utilizes sophisticated equipment to control sample or illumination beam, enabling optical sectioning of the sample. These individual sections are then stacked to reconstruct a 3D structure of the sample[55]. Additionally, single-pixel stereo vision typically involves the use of a SLM to project illumination patterns onto objects. Multiple SPDs at various locations are adopted to capture 2D images that simulate illumination from diverse directions. The 3D shape of the object can be reconstructed by comparing 2D images that exhibit precise pixel registration[56].

## 3. Single-pixel imaging based on deep learning

As the fundamental cornerstone of deep learning, deep neural networks (DNN) serve as crucial tools for simulating and solving complex problems in deep learning. This section primarily presents the principles of DNN and DL-SPI, while also providing brief explanations on the DNNs commonly used in SPI.

### 3.1 Principle of deep learning

#### 3.1.1 The fundamentals of deep neural network

DNN is a nonlinear computational model that maps input tensors to desired outputs. A standard DNN typically includes an input layer, hidden layers, and an output layer, as shown in Figure 4. Each layer is composed of numerous units, and each unit represents a simple nonlinear function. These units are connected to the units in their adjacent layers. Except for the units of the input layer, the input of each unit in the remaining layers is the weighted sum of the outputs of the units in the previous layer:

$$O_k^l = \sigma(\sum w_{kj}^{l-1} O_j^{l-1} + b) \qquad (3\text{-}1)$$

where $O_k^l$ refers to the output of the $kth$ unit in the $lth$ layer, $O_j^{l-1}$ denotes the $jth$ unit in the previous layer, $w_{kj}^{l-1}$ indicates the connection weight between the two units, and $b$ is the bias. $\sigma$ is the activation function which is adopted to increase the nonlinearity of the model[57]. The activation functions mainly include Sigmoid, ReLU, and Tanh[58]. For the input layer, the above equation is simplified as:

$$O_k^1 = V_k \qquad (3\text{-}2)$$

where $V_k$ is the input tensor.

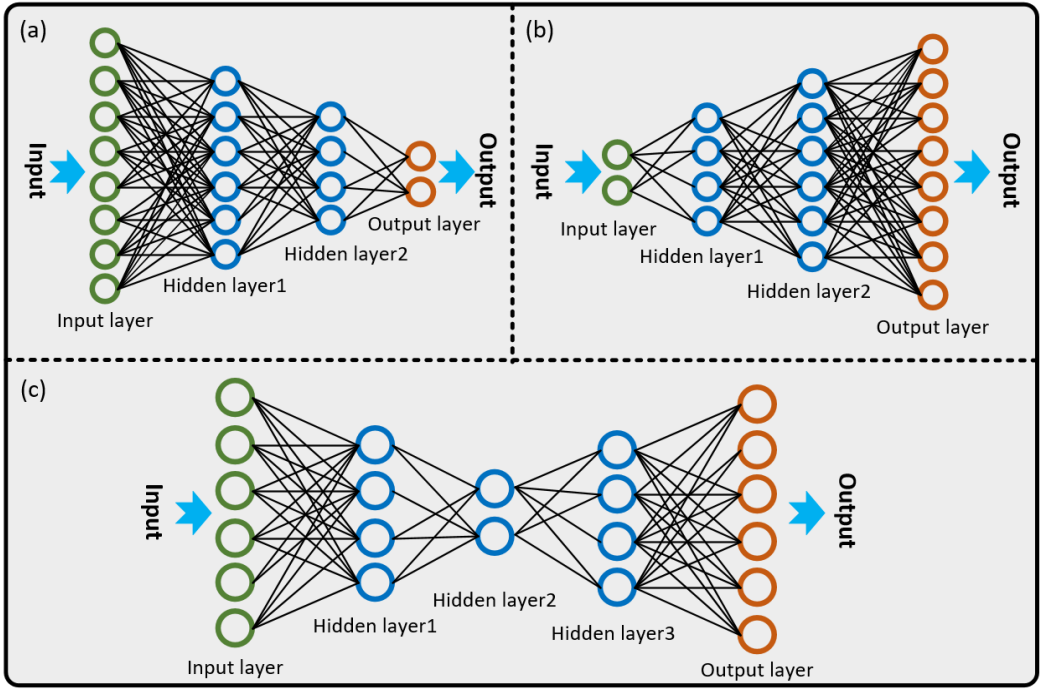

**Figure 4. Simplified schematics of DNNs with various architectures.** (**a**) Contraction Network. (**b**) Expansion Network. (**c**) AutoEncoder.

DNNs with different depths and widths are suitable for diverse tasks. The depth refers to the number of layers in DNN. The width is the number of units in each layer, which usually varies across different layers. For classification tasks where the output dimension is significantly lower than the input, a Contraction Network may be considered, as shown in Figure 4(a)[59, 60]. The up-sampling task usually generates an image with a higher resolution than the input image, and the width progressively increases toward the output, following the Expansion Network in Figure 4(b)[61]. By cascading the Contraction Network and the Expansion Network, an AutoEncoder[62] can be obtained as shown in Figure 4(c). It is suitable for image denoising[63] and image segmentation[64].

### 3.1.2 The training and testing of deep neural network

The intricate connections between simple activation units enable DNNs to handle different computational tasks. The training process aims to optimize the weights between units for the ideal connectivity, which can be divided into supervised learning and unsupervised learning[65]. In supervised learning, each training sample consists of an input and corresponding ground truth. The DNNs update their weights to generate a mapping function from the inputs to the ground truth by analyzing these training samples. Trained DNNs will be evaluated on test samples that have never been presented during training.

Suppose $N$ pairs of samples are used for training, $\{u_n, v_n\}, n = 1,2,3 \dots N$, $u$ is the input, $v$ is the ground truth. The loss function $L$ is defined as the distance between the ground truth and the actual output of all samples:

$$L = loss\_func \sum_{n=1}^{N}(DNN(u_n), v_n) \quad (3\text{-}3)$$

Where $loss\_func(\cdot)$ represents the distance metric. In general, different tasks employ different loss functions to measure the dissimilarity between predicted outputs and ground truth. Cross entropy is typically used for classification. Image denoising commonly takes the mean-square error as the loss function. Dice is usually used for image segmentation[66].

Training a DNN means minimizing the loss function by optimizing the parameters, the weights updating process of the neural network can be expressed as:

$$W_{t+1} = W_t - lr * \partial \frac{Loss}{W_t} \quad (3\text{-}4)$$

Where $W_t$ indicates the weights of the previous epoch, $W_{t+1}$ represents the updated weights, and $\partial \frac{Loss}{W_j}$ is the gradient. The $lr$ denotes the learning rate, which controls the update rate of weights. If the learning rate is too large, the weights are difficult to converge to the optimal values and the descent may oscillate around the minimum. Conversely, if the learning rate is too small, the convergence process becomes slow, and many iterations are required to achieve the desired weights. To solve this problem, a dynamic learning rate strategy is often adopted. Initially, a relatively large learning rate is set at the beginning of the training. The learning rate gradually decreases as the training progresses, which makes the model more stable at the end of training. The training process continues until a maximum number of epochs is reached or the desired accuracy is achieved.

Suppose that $M$ samples are available for testing the trained DNNs, the test error can be expressed as:

$$L = loss\_func_{test} \sum_{n=1}^{M}(DNN(u_n), v_n) \qquad (3\text{-}5)$$

The test error quantifies the generalization ability of trained DNNs. A well-generalized DNN can produce the desired outputs even for inputs that have never been seen before.

Unsupervised learning is a training strategy that does not rely on label corresponding to each input. The common unsupervised models in deep learning include autoEncoder, variational autoencoder[67], and generative adversarial network (GAN)[68]. Taking data compression and reconstruction as an example, each sample $u_n$ serves as both the input and the target output. The loss function is defined as the distance between all inputs and the actual outputs:

$$L = loss\_func \sum_{n=1}^{N}(DNN(u_n), u_n) \qquad (3\text{-}6)$$

In the training process, the DNN can extract the key features by down-sampling the inputs, and then up-sample the compressed features to recover the original inputs, achieving data dimension reduction. The weights update and testing process of the unsupervised models are identical to those outlined above, and will not be reiterated here.

**3.2 The combination of deep learning and single-pixel imaging**

The combination of deep learning and SPI can be classified into two categories: image enhancement and image reconstruction.

**3.2.1 Image enhancement through deep neural network**

Image enhancement refers to improving the quality of rough images using deep learning models, as shown in Figure 5(a). The supervised strategy of image enhancement can be broadly described as follows[69]: (1) Construct a dataset consisting of pairs of high-quality images and reconstructed images. The high-quality images can be acquired by sufficiently accurate imaging devices, or by utilizing publicly available datasets, such as ImageNet[70]. The reconstructed images refer to rough images generated by SPI or adding noise to the high-quality images. (2) Develop a DNN and train it using the dataset constructed in the previous step. (3) Evaluate the performance of the trained DNN by inputting reconstructed images to obtain superior outputs.

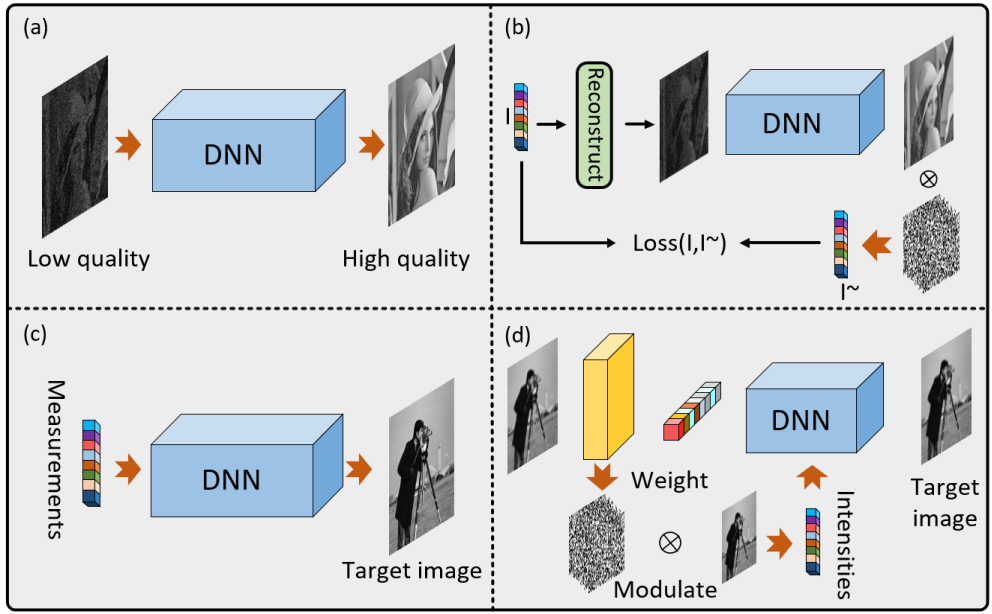

**Figure 5. The different categories of DL-SPI.** (**a**) Image enhancement. (**b**) Unsupervised image enhancement. (**c**) Image reconstruction. (**d**) Joint optimization.

The unsupervised image enhancement strategy does not require training the DNN[35]. This approach initially employs traditional algorithms to reconstruct rough images in Figure 5(b). Subsequently, these rough images are inputted into the DNN to reconstruct high-quality images. Then, the estimated 1D measurements of a high-quality image are obtained by SPI forward model, which involves performing a weighted summation between the patterns and target images. Finally, the weights of the DNN are updated to minimize the error between the real and estimated measurements.

**3.2.2 Image reconstruction through deep neural network**

Image reconstruction refers to directly restoring the images from the 1D measurements through neural networks, as shown in Figure 5(c). The supervised strategy of image reconstruction also includes dataset preparation, model design, model training, and model evaluation[71-73]. Different from image enhancement, the dataset for image reconstruction consists of pairs of 1D measurements and ground truth. The 1D measurements can be collected by bucket detector or generated by a SPI forward model. The unsupervised strategy for image reconstruction is similar to that for image enhancement, except that DNNs directly decode high-quality images from noise latent vectors or 1D measurements instead of low-quality images[74, 75].

In addition, DNNs can be utilized to generate optimal patterns in image reconstruction to make the sampling high-efficiency and hardware-friendly. This approach sets the modulation patterns as trainable parameters of DNN and jointly optimizes them with the subsequent reconstruction network in Figure 5(d). After training, the parameters of the neural network layer represent the optimal patterns that match the reconstruction network[76].

**3.3 Deep learning algorithms commonly used in single-pixel imaging**

**3.3.1 Convolutional Neural Network**

Convolutional neural network (CNN), introduced by Le et al., is a deep learning model designed for processing gridded data. A typical CNN for image reconstruction is depicted in Figure 6(a)[77]. It usually contains convolutional layers, pooling layers, activation layers, and transposed convolution layers. The convolutional layer is implemented by calculating the dot product between the convolution kernel and the feature maps. The pooling layers down-sample the feature maps. The activation layers serve to enhance the nonlinearity of CNN. The transposed convolution layers up-sample the feature maps. The weight refinement of CNN is accomplished by backpropagation algorithm. The convolution layers allow the CNN to capture spatial features from the input data, ranging from low-level textures to high-level shapes. CNNs commonly employed in SPI include DenseNet[78], ResNet[79], Efficient-net[80], U-net[81].

**3.3.2 Generative Adversarial Network**

GAN is a special type of DNN consisting of a generator and a discriminator, which structure is depicted in Figure 6(b). The generator is responsible for generating excellent images that can deceive the discriminator. The discriminator attempts to distinguish between the ground truth and fake images generated by the generator. The generator and discriminator are updated alternately during the training process. When the generator is being trained, the parameters of the discriminator are kept fixed, and vice versa. The training process continues until a state of equilibrium is achieved. The trained generator can transform a 1D vector into a 2D image. In SPI, several variants of GAN are often used to recover images, such as conditional generative adversarial networks[82], and deep convolutional generative adversarial networks[83].

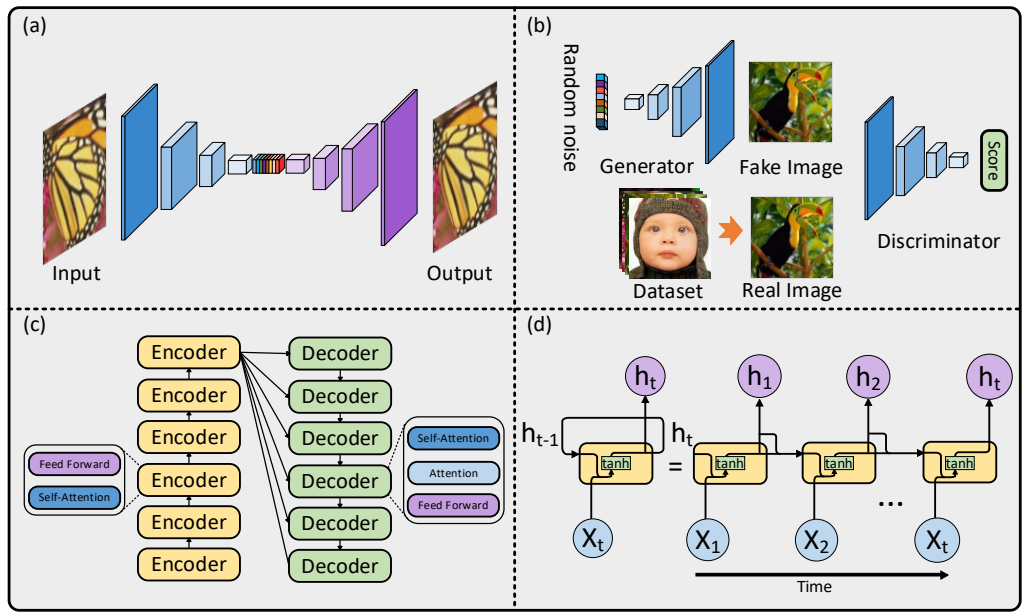

**Figure 6. The DNNs commonly used in SPI.** (**a**) Convolutional Neural Network. (**b**) Generative Adversarial Network. (**c**) Transformer. (**d**) Recurrent Neural Network.

**3.3.3 Transformer**

The transformer is essentially a specific AutoEncoder, which consists of six encoders and six decoders[84], as shown in Figure 6(c). Each encoder includes two sub-layers, a multi-Head Self-Attention layer and a Position-wise Feed Forward layer.

Similarly, each decoder contains three sub-layers, a Multi-Head Self-Attention layer, an Attention layer, and a Position-wise Feed Forward layer. The input to each encoder is the output of the previous encoder, while each decoder receives input from the previous decoder as well as the remaining encoders. The encoding block maps the input sequence into encoded information, while the decoding block converts the encoded information into the desired output. The multi-head attention mechanism can capture correlations between different vectors regardless of their spatial distance, which allows the model to discern relationships and nuances between the vectors. In SPI, the transformer is typically used to directly process 1D intensity[34].

### 3.3.4 Recurrent Neural Network

Recurrent neural network (RNN) is a kind of neural network specifically designed for processing serialized data[85]. The fundamental structure of RNN consists of recurrent units, as depicted on the left side in the Figure 6(d). The RNN stores the output of the current epoch in a storage unit, which is then used as input along with the input data when the next epoch is entered. This process can be viewed as multiple assignments of the same neural network at different time steps, with each neural unit passing the message to the next. The expanded structure of the RNN is shown on the right side in Figure 6(d). In image processing tasks, 2D image data are usually input into the RNN pixel by pixel. Similar to transformer, RNNs can generalize long sequences to get the desired result. Thus, RNNs offer significant advantages in directly transforming intensities into images[86].

## 4. Single-pixel imaging realization with deep learning

DL-SPI has demonstrated significant potential in various fields. This section explicates the research progress of DL-SPI in super-resolution, imaging through scattering media, imaging at the single-photon level, optical encryption, color SPI, and image-free sensing.

### 4.1 Super-resolution imaging

The spatial resolution of an optical imaging system is a quantification of its ability to distinguish fine details and is a crucial metric for evaluating image quality. The resolution is primarily constrained by discrete sampling of the detectors and diffraction of the optical system. The former is referred to as image resolution, indicating that the discrete sampling of low-resolution detectors may result in the loss of image details. The latter is known as optical resolution, describing the overlap and blurring of neighboring target points caused by diffraction [87, 88].

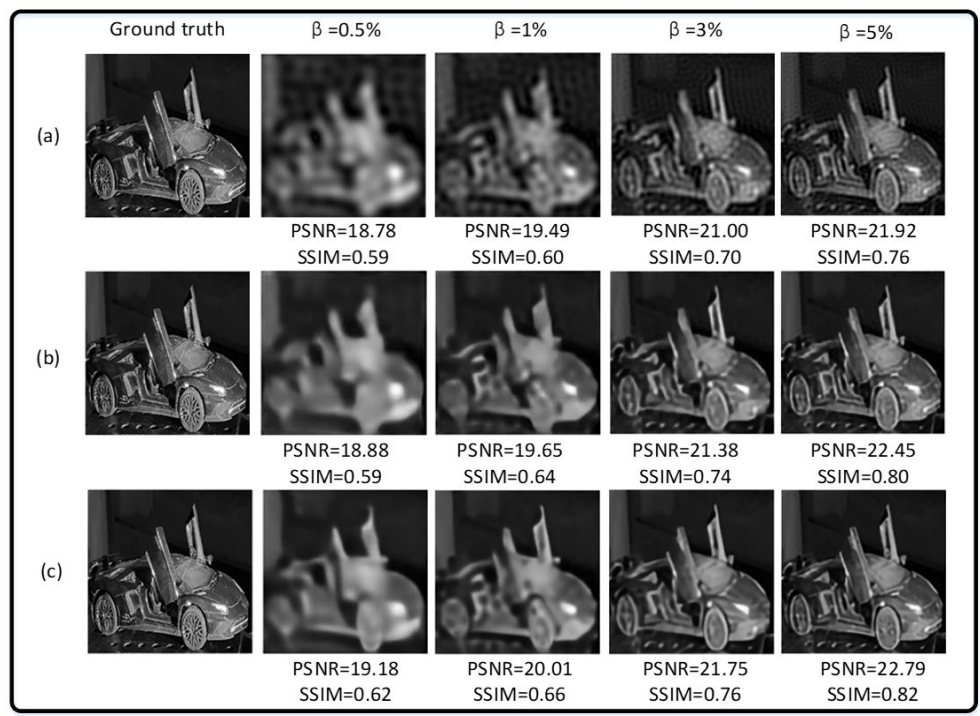

**Figure 7.** Comparison of super-resolution images generated by different algorithms[36]. (**a**) Traditional Fourier SPI[22]. (**b**) Deep convolutional Fourier SPI[89]. (**c**) Super-resolution Fourier SPI. The sampling rates ranged from 0.5% to 5%, respectively.

In terms of improving image resolution, DNNs have been extensively applied to improve image resolution by up-sampling. For instance, researchers utilized a super-resolution network that integrates GAN and U-net to improve the resolution of Fourier SPI. As shown in Figure 7, when the sampling rate is less than 1%, the reconstruction results generated by traditional Fourier SPI (Figure 7a) and deep convolutional Fourier SPI (Figure 7b) have significant ringing effects. In contrast, the proposed super-resolution network effectively mitigates the impact of the ringing and successfully raises the resolution of reconstructed images from 128×128 to 256×256 in Figure 7c[36].

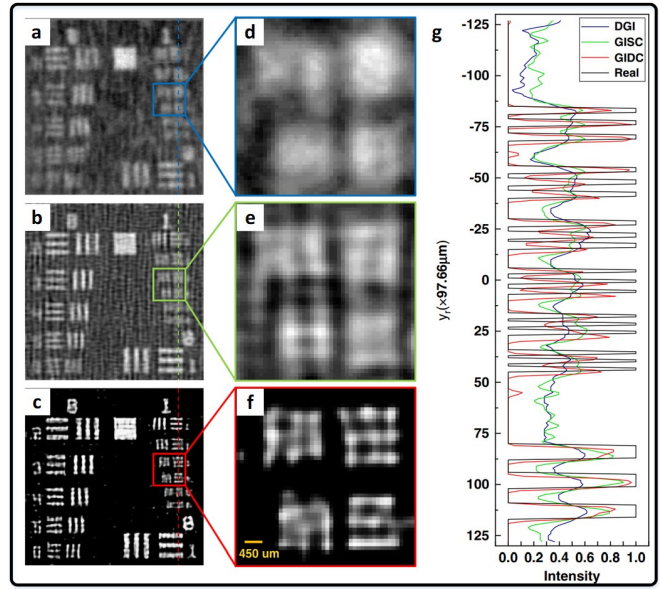

**Figure 8.** Experimental results for USAF resolution target using various algorithms[35]. **(a-c)** Reconstructed images are obtained by differential ghost imaging (DGI)[90], compressive sensing ghosting imaging (GICS) and unsupervised U-net (GIDC), respectively. **(d-f)** Zoom-in images of different ROIs of (a-c). **(g)** Normalized pixel values of the dashed line in (a-c).

In addition to enhancing the resolution of images, DNNs are also frequently used to tackle the challenges of noise and blurring caused by limitations of the optical system for optimizing optical resolution[32, 91-95]. Recently, Wang et al. proposed an unsupervised image enhancement framework, which demonstrates the potential to surpass diffraction limits and exhibits superior performance in terms of both linewidth and sharpness of reconstructed images. Figure 8 illustrates the imaging results of the USAF resolution target using various algorithms. By analyzing the linewidth of the reconstructed images, the diffraction limit of the optical system is reduced from 683.59 μm to 353.55 μm, resulting in a resolution increase of about 2 times[35]. Furthermore, Li et al. also applied an untrained U-net in multimode fiber imaging to improve the resolution. The evolution of image estimates along the feature sizes is shown in Figure 9. The smallest features of samples 1-4 are 1.4, 1.8, 2.5, and 4 times smaller than the diffraction limit (65 μm), but the DNN can still provide sub-diffraction spatial resolution of the multimode fiber imaging system[96].

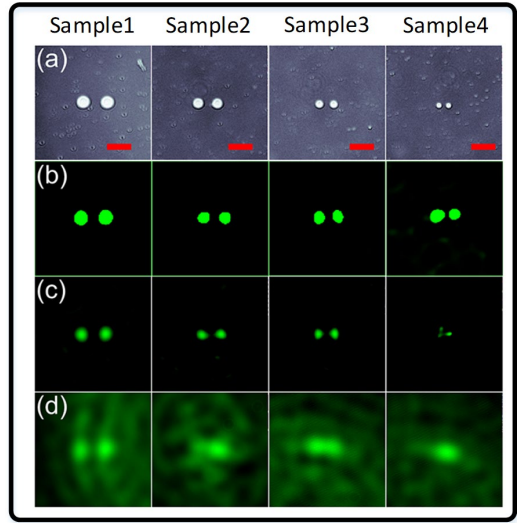

**Figure 9**. **The evolution of image estimates along the feature sizes in multimode fiber imaging**[96]. (**a**) Four samples with different feature sizes (43 μm, 36 μm, 26 μm, 16 μm). (**b-d**) Reconstruction results from the same dataset using the untrained U-net, basis pursuit algorithm, and GI algorithm.

**Table 1. Summary of deep learning algorithms for enhancing optical resolution.**

| Method | Resolution | Sampling rate | Real-time imaging | Running Time (s) |
|---|---|---|---|---|
| CGANCGI[97] | 28×28 | 8% | - | - |
| AECGI[98] | 50×50 | 8% | - | 0.04 |
| TransUnet[99] | 64×64 | 30% | - | 0.004 |
| Res-AutoEnconder[100] | 64×64 | 6.25% | ≥80 Hz | - |
| DeepGhost[101] | 96×96 | 20% | 4-5 Hz | 0.13 (DGI+DNN) |
| DDANet[32] | 128×128 | 3.5% | - | - |
| DAttNet[102] | 128×128 | 5.45% | - | - |
| DCAN[103] | 128×128 | 2% | 30 Hz | - |
| U-net[104] | 128×128 | 30% | - | 0.005 |
| SR-FSI[36] | 256×256 | 1% | - | 0.0327 |
| CDAE [105] | 256×256 | 3.7% | - | - |

Additionally, in the term of optimizing optical resolution, researchers have also devoted themselves to developing deep learning algorithms capable of reconstructing images from extremely sparse sampling rates, the summary of these algorithms is provided in Table 1[97-99, 104]. For simple target images, the existing deep learning algorithms can reconstruct the high-quality images at the sampling rates of around 3-5% or even lower[32, 36, 102, 105]. These achievements significantly reduce the reconstruction time for SPI. Moreover, most of the deep learning algorithms exhibit running times below 100 ms, and even as low as 10 ms, which offers a potential solution for achieving real-time SPI[100, 101, 103]. The excellent deep learning algorithms can reconstruct videos with a resolution of 128×128 / 64×64 at a frame rate of 30[103] / 80[100] per second.

**4.2 Imaging through scattering media**

Imaging through scattering media is a difficult and unsolved problem, which holds a vital position in various fields including remote sensing[106], underwater imaging[107], and biomedical imaging[108]. Perturbations (scattering, refraction, or absorption) occurring within the complex medium hinder the propagation of light along straight paths, leading to distorted or blurred imaging results[109]. The distinctive imaging mode of SPI offers significant advantages in imaging through scattering media[110]. Presently, studies in this domain primarily concentrate on enhancing the quality of reconstructed images through advanced reconstruction algorithms. Deep learning algorithms present an especially appealing option owing to their powerful image-processing ability, enabling the SPI to effectively operate in more challenging environments.

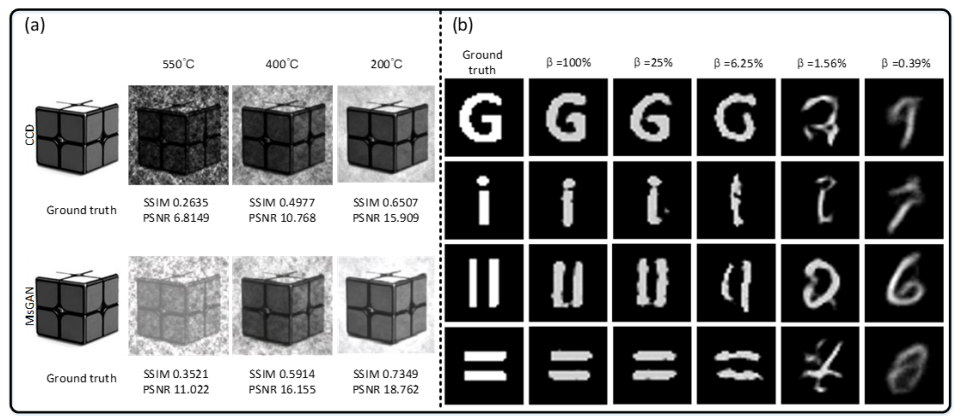

**Figure 10. Deep learning applied to turbulence-immune imaging.** (**a**) Experimental results are obtained by CCD camera and multi-scale GAN (MsGAN), respectively. 550 °C, 400 °C, and 200 °C represent levels of turbulence, ranging from strong to weak in descending order[38]. (**b**) English alphabets and double-seam patterns reconstructed by the deep learning algorithm that combines U-net and ResNet. The sampling rates range from 100% to 0.39%.

Turbulence-immune imaging is a classical challenge in the field of imaging through scattering weak media. DL-SPI has exhibited remarkable efficacy in this domain. For example, Zhang et al. proposed a multi-scale GAN (MsGAN) to clear the low-quality images obtained by computational ghost imaging. The experimental results acquired by the CCD camera and MsGAN are illustrated in Figure 10(a). In strong turbulence (550 °C), the Structural Similarity (SSIM)[111] and Peak Signal-to-Noise Ratio (PSNR)[112] of images recovered by MsGAN are 33.6% and 61.7% higher than those of turbulent images captured by a CCD camera. In weak turbulence (200 °C), the MsGAN achieves an average improvement of 12.9% in SSIM and 17.9% in PSNR[38]. Furthermore, the deep learning algorithms can also be directly applied to nonlinear inverse problems. The target is modulated by a set of random patterns that have undergone Fresnel diffraction over three meters or more. The imaging results are reconstructed by a deep learning algorithm that combines U-net and ResNet in Figure 10(b) [113]. The network demonstrates its effectiveness in reconstructing images of English alphabets and double-seam patterns when the sampling ratio β exceeds 25%, despite being trained on MNIST dataset.

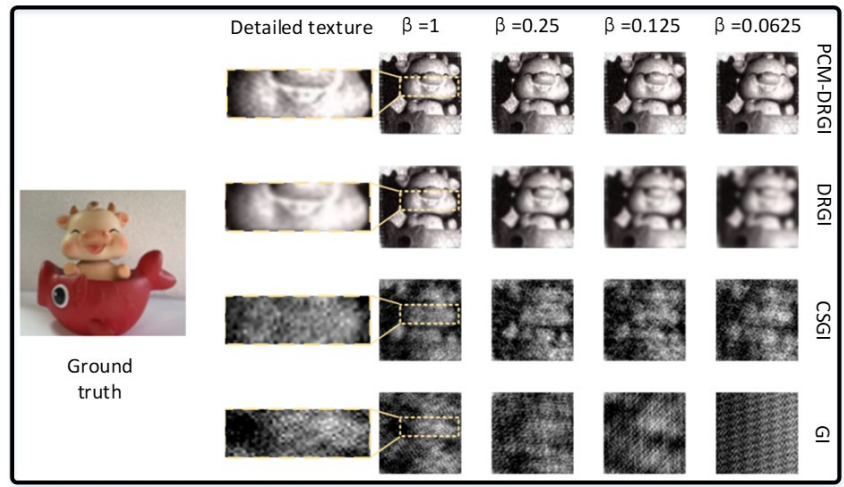

**Figure 11. Recovery ability comparison of various algorithms under diffuser scattering**[37]**.** The algorithms adopted are degradation-guided ghost imaging (DRGI) based on PCM (PCM-DRGI), DRGI, compressive sensing ghosting imaging (CSGI), and GI. Sampling rate β = 1, 0.25, 0.125, 0.0625, respectively.

For imaging through strong scattering media, a lot of studies have indicated that the weight parameters of DNN, ranging from millions to tens of millions, can perfectly compensate for the complex nonlinear processes of scattering and absorption[114]. Similar to turbulence-immune imaging, several studies adopt the image enhancement technique to refine the quality of images reconstructed by traditional algorithms[115]. Other studies employ the image reconstruction technique to transform the scattered intensities into high-quality images[116-118]. It is noteworthy that the introduction of prior knowledge often provides more effective constraints on the convergence of DNNs. Gao et al. utilized photon contribution model (PCM) data to guide the DNN in learning the mapping operator, which converts degraded images into clear images in the Narasimhan atmospheric scattering model[119]. This operator facilitates deliberate guidance of information flow within the reconstruction sub-network, thereby enabling the weight parameters with the capability for automatic feature extraction, and the results are showed in Figure 11. The proposed PCM-DRGI algorithm demonstrates superior performance than other algorithms, especially in the low sampling rates (6.25%)[37].

**4.3 Imaging at the single-photon level**

In extreme environments or long-distance scenarios, the echo signal from the targets is extremely faint, even reaching the single-photon level. It is difficult for traditional imaging techniques to obtain enough intensity information to present a high-quality image. In contrast, single-photon imaging offers remarkable sensitivity at the photon level, making it an optimal solution for imaging in extreme or long-distance scenes[47]. Currently, although the single photon avalanche detector array cameras have achieved remarkable development, they still face the challenge of inconsistent pixel parameters and high dark count rate. However, the SPI with a SPD (photon-level SPI) avoids these challenges and can produce more uniform and stable images[120]. Therefore, photon-level SPI has attracted considerable attention, including the exploration of deep learning applications in this domain.

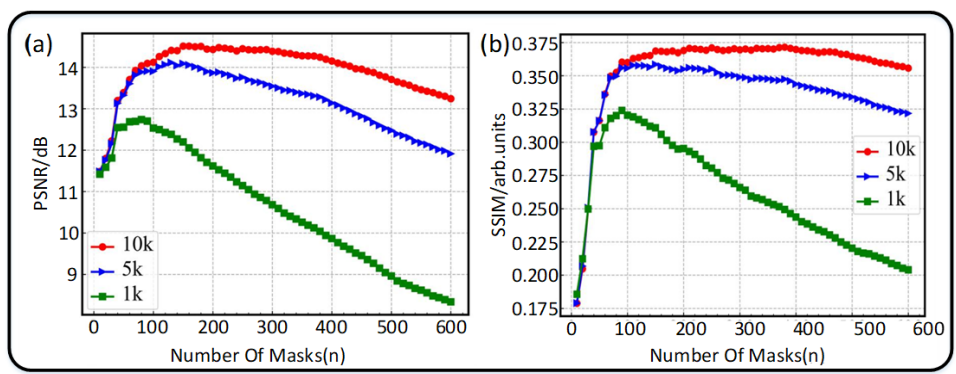

**Figure 12. The relationship among quantum shot noise, the number of masks and image quality**[121]. (**a**) The variation of PSNR along with the number of masks under different photon counts. (**b**) The variation of SSIM along with the number of masks under different photon counts.

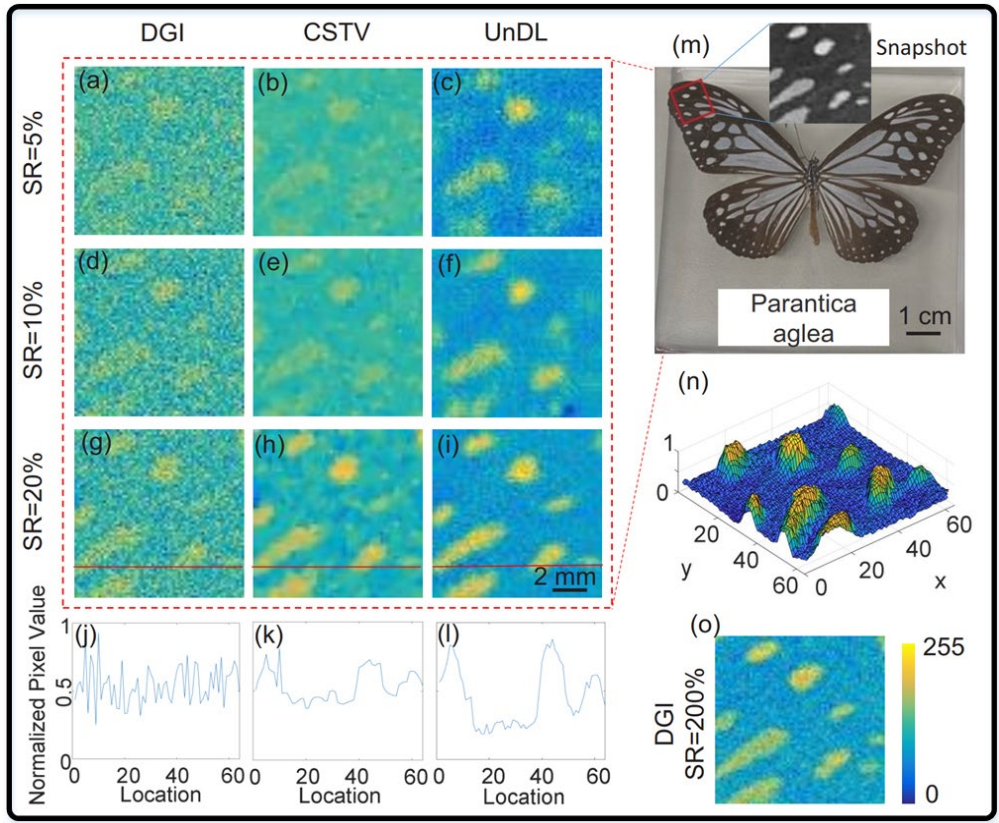

**Figure 13. Retrieval of a butterfly specimen[92].** (**a-i**) Imaging results of DGI, compressive sensing total variation regularization (CSTV)[122], and anti-noise deep learning algorithm (UnDL). (**j-l**) The cross-sectional profile of the red line in (j-l). (**m**) Picture of the butterfly specimen and a snapshot of ROI. (**n**) The contour map of (i). (**o**) DGI result with a sampling ratio of 200%. The imaging quality of all methods improves as the sampling rate increases.

Since the high sensitivity of photon-level SPI to noise, the traditional reconstruction algorithms may encounter difficulties in restoring clear images when the echo signal is extremely weak or the environmental noise is pronounced. In such cases, increasing the sampling rate to enhance the signal-to-noise ratio of images is ineffective. Fan et al. have demonstrated that the additional measurements may introduce more noise in extreme scenarios. The relationship between among quantum shot noise, the number of masks and image quality is shown in Figure 12[121]. The image quality initially improves and subsequently gets worse with an increasing number of patterns when the photon count rate is low. Fortunately, deep learning algorithms have emerged as a potent solution, harnessing their robust analytical and learning abilities[123, 124]. Figure 13 presents the results of a novel anti-noise framework. This deep learning framework utilizes the surrounding pixel values to accurately predict each pixel.

Compared with other deep learning algorithms which only rely on individual pixel information, this framework has shown superior anti-noise capabilities[92]. The cross-sectional profile of the red line in Figure 13 shows that the proposed method has more obvious peaks, substantiating its efficacy. Additionally, considering the potential impact of time-varying interference on image fidelity, Jia et al. proposed an unsupervised data-domain translation technique, inspired by the generalized domain translation technique[125]. This deep learning algorithm integrates a feature extraction module with a denoising module to reconstruct images, achieving 10 photons per pixel or even less.

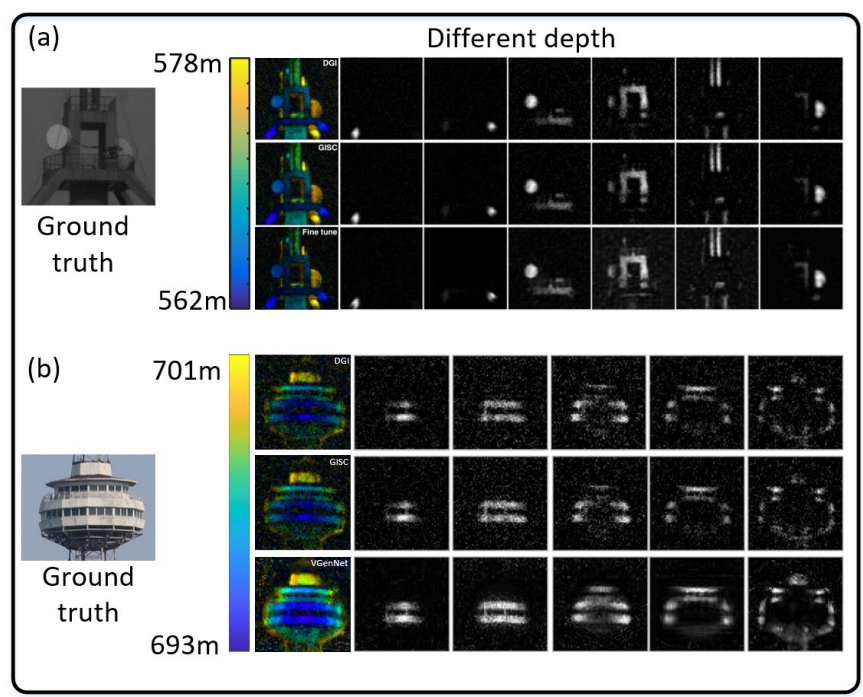

Figure 14. The 2D depth slices and 3D images of the different TV towers. (**a**) TV tower at distances of 570 m obtained by DGI, ghost imaging via sparsity constraint (GISC), and physics-enhanced deep learning algorithm (Fine tune)[4]. (**b**) TV tower at distances of 693 m recovered by DGI, GISC, and VGenNet[40].

The deep learning algorithm has also shown its efficiency in single-pixel LiDAR. Radwell et al. first applied the DNN to improve the imaging quality of single-pixel LiDAR. They successfully reconstructed 3D images of an object located 28 m away[126]. Since then, more and more researchers have endeavored to develop innovative deep learning algorithms for the retrieval of 3D images over long distances. For example, Figure 14 (a-b) illustrate the 3D images of the TV tower at distances of 570 m[4] and

693 m[40], respectively. The end-to-end deep learning models adopted are physics-enhanced network and variable generative prior enhanced network (VGenNet) respectively. The incorporation of physics not only optimizes the illumination patterns but also improves the generalization ability of the framework. The images reconstructed by deep learning methods exhibit superior quality in terms of sharpness, noise reduction, and contrast. The VGenNet ensures the interpretability and universality of the deep learning model, which paves the way for its practical application.

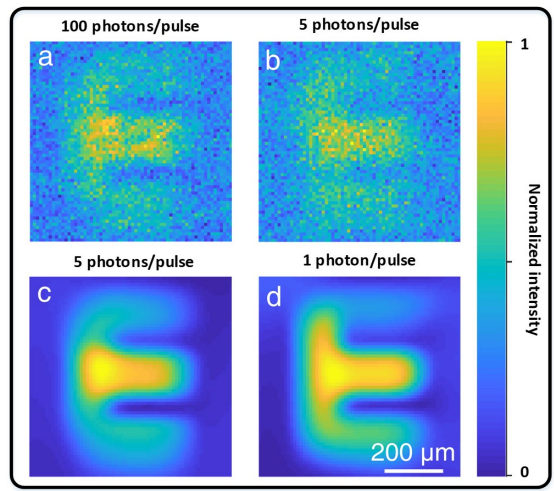

**Figure 15. Deep learning applied to improve the quality of mid-infrared images**[7]. (**a-b**) Images reconstructed by CS algorithm for Mid-infrared illumination intensity of 100 and 5 photons/pulse, respectively. (**c-d**) Images processed by plug-and-play algorithm[127] under Mid-infrared illumination intensity of 5 and 1 photons/pulse, respectively. The plug-and-play de-noising algorithm can enhance the image quality in both cases of 5 and 1 photons/pulse.

For invisible wavelength photon-level SPI, the beam steering performance of SLM is limited by diffraction effects at invisible wavelengths, failing the spatial modulation[2]. To avoid this limitation, Wang et al. proposed a nonlinear frequency-conversion process to promote mid-infrared optical spatial modulation. They adopted a plug-and-play de-noising algorithm to suppress the noise existing in low-quality images. The deep learning algorithm can generate high-quality images in the condition of extremely low flux with 1 photon/pulse under a sample ratio of 25% (Figure 15) [7].

**4.4 Optical encryption**

Optical encryption is an information encryption technology that utilizes the abundant degrees of freedom of light to encode and decode target information, such as amplitude, phase, frequency, and polarization[128-130]. Recently, SPI has rapidly emerged as an alternative optical encryption approach owing to the indirect imaging modality. Various SPI encryption schemes have been developed to enhance the security of optical encryption systems, including color image encryption[131], steganography techniques[132], and visual cryptography encryption[133]. In these cryptosystems, the sender and receiver transmit a large number of patterns and corresponding intensities detected by the bucket detector, which serve as the key and ciphertext respectively. However, a lot of patterns not only easily attract the attention of eavesdroppers but also increase the burden of transmission. Therefore, the deep learning algorithm is commonly perceived as a more robust and secure technique in SPI optical encryption. It can effectively enhance the security of the system while achieving data compression at the same time[134, 135].

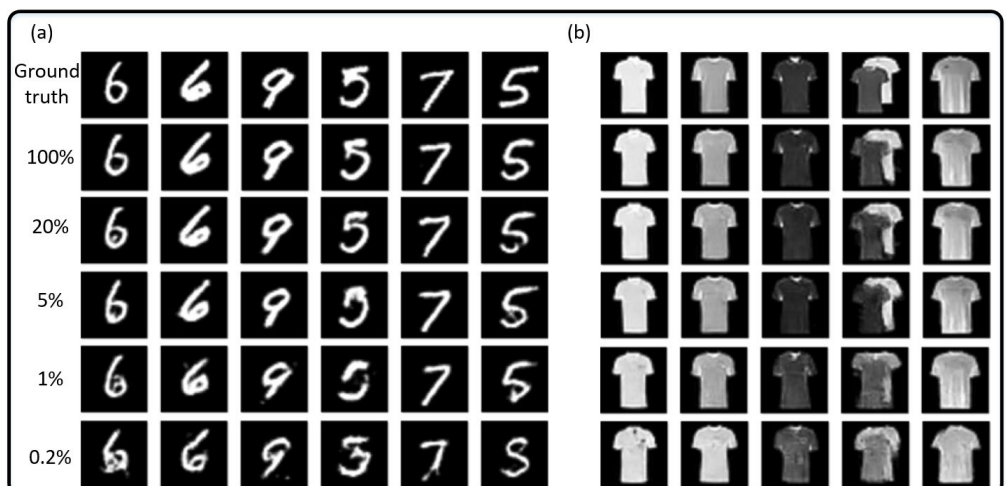

**Figure 16. Retrieved object images at different sampling rates**[41]. (**a**) MNIST test data. (**b**) fashion-MNIST test data.

End-to-end deep learning algorithms are trained to directly reconstruct images from measurements. For instance, an improved conditional generative adversarial networks is proposed to directly decode the encrypted intensities and reconstruct the hidden images, and the experimental results are shown in Figure 16[41]. Even when the sampling rate is as low as 0.2%, the enhanced conditional generative adversarial networks can restore clear images. However, this decryption approach is susceptible to chosen-plaintext attacks, because the attackers can exploit a limited set of plaintext-ciphertext pairs to regress the forged key suitable for the SPI cryptosystem[136]. Therefore, SPI optical encryption systems often incorporate additional encryption technologies with deep learning algorithms to enhance security[137-139]. Wang et al. encrypted the binary compressed measurements using orthogonal codes. Multiple pieces of encrypted data streams can be combined and transmitted without interference. The receiver transforms the binary measurements to images via a DenseNet, as shown in Figure 17. This scheme can not only compress data during transmission but also allow concurrent access for multiple users[138].

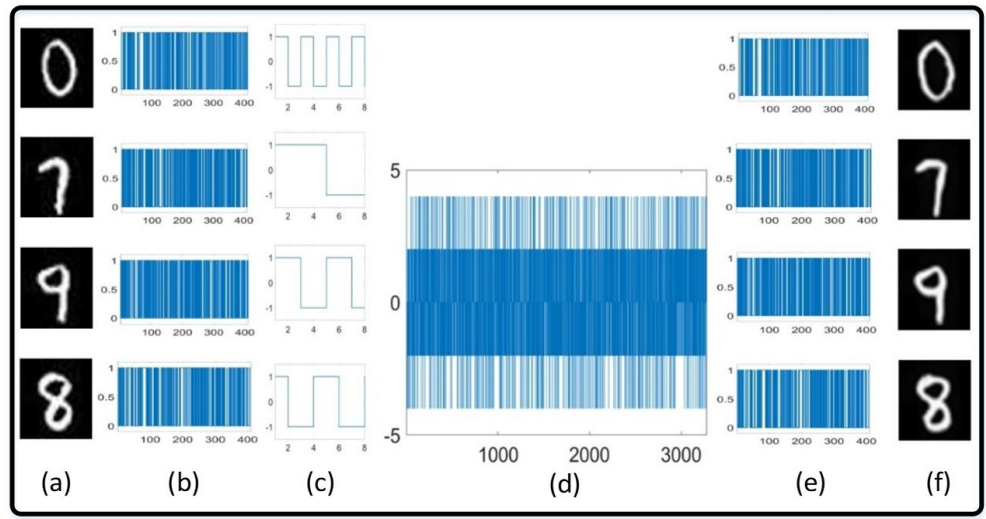

**Figure 17. The process of SPI Optical encryption based on orthogonal coding**[138]. (**a**) Plaintext images. (**b**) Binary intensities. (**c**) Orthogonal code sequences. (**d**) Ciphertext. (**e**) Recovered binary intensities. (**f**) Reconstructed images.

**4.5 Color single-pixel imaging**

Color SPI can achieve independent reconstruction of red, green, and blue channels by frequency-division multiplexing[140], single-time measurement with multiple detectors[141] or multiple-time measurement with a single detector. Compared with the gray SPI, the complex color SPI system may require longer imaging times, and the unknown color response coefficient can inevitably lead to color distortion. Different from the full channels sampling methods, the color-filter-array-based SPI realizes color image reconstruction without altering the structure of gray SPI. This approach utilizes a color filter array to modulate patterns and adopts the appropriate algorithms, such as deep learning algorithms, to recover color images[142]. Recently, the deep learning algorithms have been introduced into these strategies, which can significantly mitigate the complexity of the system and reduce the imaging time.

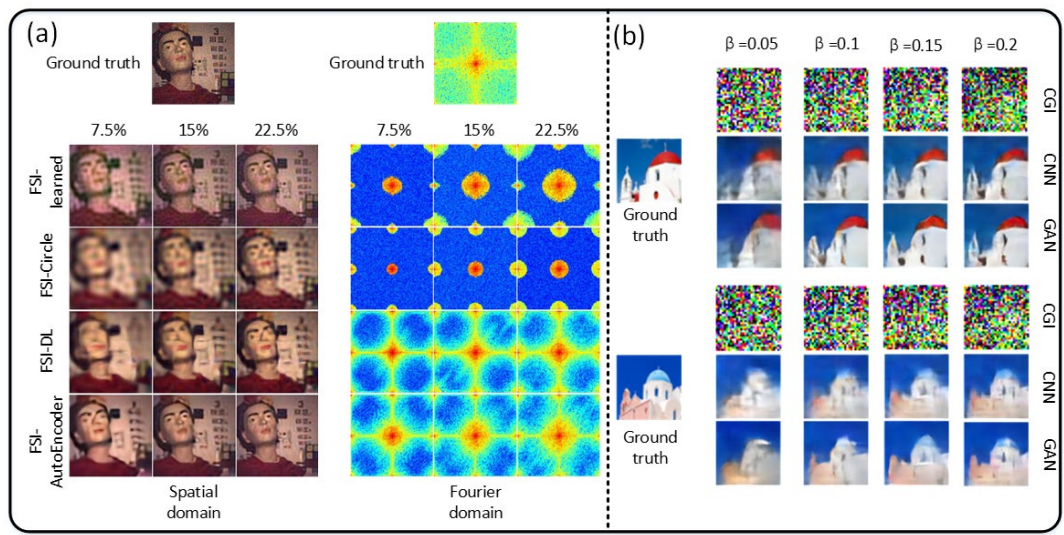

**Figure 18. Color imaging implemented by DL-SPI.** (**a**) Experimental results of spatial domain and Fourier domain of different reconstruction methods with various sampling rates[42]. (**b**) The results are optimized by different algorithms with various sampling rates[143].

For the color SPI based on frequency-division multiplexing, Jiang et al offered a multi-channel deep convolutional autoencoder network to process multi-channel bands[144]. Compared to traditional reconstruction methods, the network achieved large advancements in both the image quality and reconstruction speed, successfully realizing hyperspectral single pixel video reconstruction at a rate of 12 fps.

In the color-filter-array-based SPI, deep learning algorithms are commonly used to colorize and demosaic gray mosaic images recovered by traditional algorithms. The deep learning models can directly reconstruct high-definition color images from gray mosaic images[145] or can perform colorization and demosaicking operations at multiple stages[146]. For example, Huang et al. propose a Fourier single-pixel imaging based on color filter array. They employ an autoencoder to convert gray images into colored images, even with the sampling ratio as low as 7.5% (Figure 18(a))[42]. In addition, the deep learning algorithm can also directly optimize noise color images based on random patterns. The results with various sampling rates are shown in Figure 18(b). As anticipated, the deep learning models can greatly improve the color images reconstructed by computational ghost imaging and the proposed GAN performs best in these methods [143].

## 4.6 Image-free sensing

From a technical perspective, the ultimate goal of SPI is not limited to recovering high-quality images in practical applications. The more important task of SPI is to provide images for advanced computer vision tasks, such as object classification, segmentation, and detection. However, these computer vision tasks typically only use partial features of the image, rather than the whole image. Therefore, images generated by SPI may contain a large amount of redundant information[147]. The acquisition, decoding, transmission, and storage of these data incur additional hardware and computational costs. Thus, advanced sensing tasks can be performed directly from the limited number of 1D measurements without reconstructing images, as shown in Figure 19.

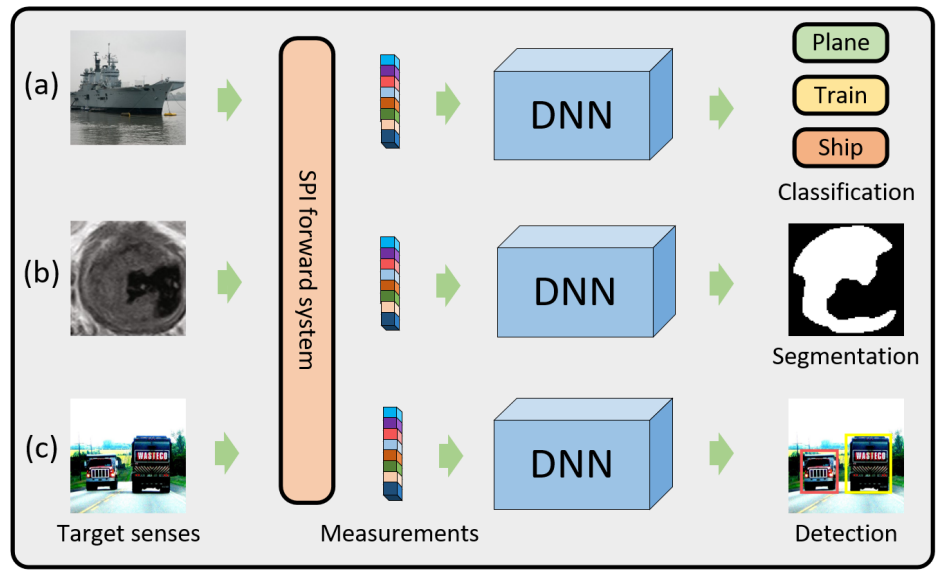

**Figure 19. The principle and different tasks of image-free sensing**[70]. (**a**) Image-free classification. (**b**) Image-free segmentation. (**c**) Image-free detection.

Image-free sensing technology is proposed in 2018 by S. Ota et al. They used the pre-trained support vector machine[148] to classify the continuous fluorescence signals generated by the cells for identifying different cell types. The detailed process is shown in Figure 20(a)[43]. Since then, a continuous stream of classification tasks based on image-free sensing has emerged, the detailed results are shown in Table 2[149-154]. For the classification of static objects, almost all image-free sensing models based on deep learning can achieve an average classification accuracy exceeding 90% even at low sampling rates. In particular, a deep learning framework based on Efficient Net achieves remarkable accuracy, reaching up to 96% in handwritten digit classification tasks, even with a low sampling rate of only 3%[152]. For classifying complex moving objects, image-free sensing technology can effectively mitigate the impact of motion blur on classification accuracy. Zhang et al. successfully identified the objects moving at a velocity of 3.61 m/s within the field-of-view of 45 mm×45 mm. It is worth noting that the classification framework is very simple, consisting of only one deconvolution layer and three fully connected layers[154].

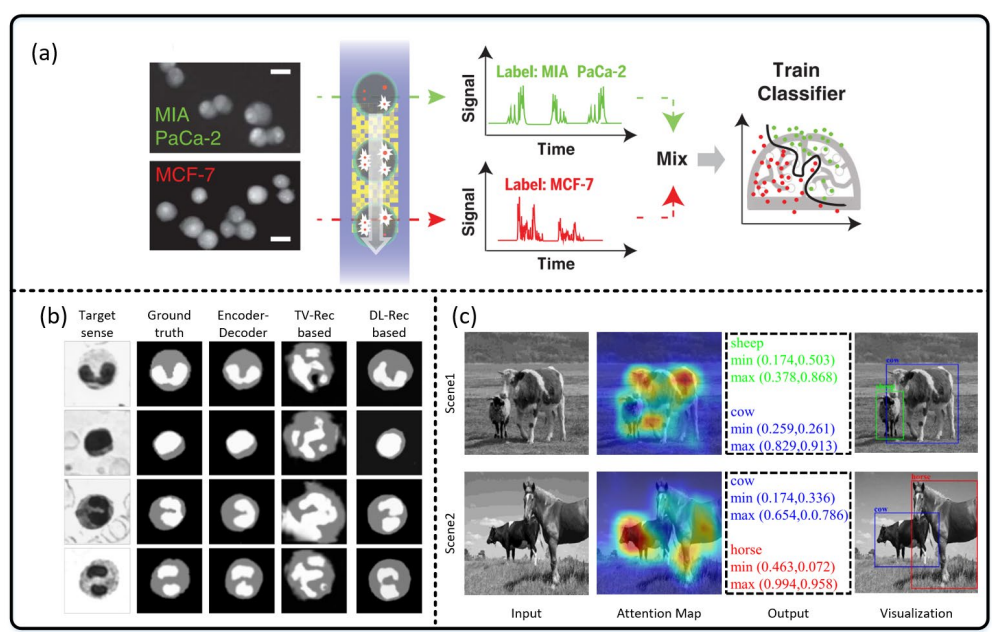

**Figure 20. Representative experimental results for different image-free sensing tasks.** (**a**) The process of training a support vector machine in ghost cytometry[43]. (**b**) Segmentation results of the encoder-decoder model, CSTV, and deep-learning-based reconstruction method[103] under a 1% sampling ratio[44]. (**c**) Detection results of the deep learning model under various challenging conditions with a sampling rate of 5%[45].

**Table 2. Summary of diverse deep learning algorithms for image-free sensing.**

| Methods | Sampling rate | MNIST | Fashion-MNIST | Multi-Character |
|---|---|---|---|---|
| DNN[149] | 12.67% | 91% | - | - |
| SPDH[150] | 25% | 81.39% | - | - |
| SPS[151] | 3% | 96.68% | 84.40% | - |
| FCFNN[152] | 3.13% | 94.50% | 84.48% | - |
| CRNN[153] | 5% | - | - | 87.65% |
| CNN[154] | 0.1% | 100% (3.61 m/s) | - | - |

The image-free sensing technology has also demonstrated its efficacy in image segmentation tasks[44]. As shown in Figure 20(b), an autoencoder is trained to infer segmentation maps from 1D measurements. The pixel accuracy of the deep learning model reaches above 96% even at a sampling rate as low as 1%. Moreover, Peng et al. first applied image-free sensing technology to obtain the location of objects for the object detection task. They proposed a deep learning model based on the Transformer architecture and multi-scale Swin-Conv residual blocks, which can simultaneously extract the category, location, and size of all objects existing in the scene. The model achieves a detection accuracy of 82.41% at a sampling rate of 5% with a refresh rate of 6.3 fps ( Figure 20(c)) [45].

**5. Conclusion and Outlook**

This review focuses on the latest advancements in DL-SPI and related applications. In the evolution from its first implementation to reconstruct 256 × 256 images with sampling rates as low as 5% or even lower, DL-SPI has showcased remarkable advancements in both imaging quality and speed. Furthermore, the compatibility of deep learning with SPI empowers the image-free sensing to efficiently complete advanced computer vision tasks in a cost-effective manner, achieving accuracy exceeding 80%. However, the wider application of DL-SPI within the imaging community still constrained by several factors, such as dataset dependence, algorithm versatility, minimal physical limitations, and the incorporation of prior knowledge.

Consequently, the primary focus in DL-SPI lies in the design of more efficient and generalizable deep learning algorithms for future directions[155]. Primarily, considering the challenges and high costs associated with dataset acquisition in SPI, it is imperative for deep learning algorithms to reduce their reliance on extensive labeled data. Approaches such as unsupervised learning, data augmentation[156], transfer learning[157], and few-shot learning[158] hold promise in effectively addressing these challenges. Furthermore, there is a pressing need to enhance the interpretability and fairness of deep learning algorithms. Integrating prior knowledge and physical constraints into DNNs offers a promising avenue. This not only increases the interpretability of DNNs but also restricts their learning space, ultimately improving the performance of deep learning algorithms.

Additionally, it is worth exploring deep optics strategies that leverage deep learning algorithms to optimize optical design and expand ability of information transmission[159, 160]. Deep learning can be applied to designing various components in imaging systems. For instance, DNNs can be utilized to investigate structured light fields[161] that align with current imaging requirements, allowing for the selection of novel light sources (super-bunched source[162], polarization entanglement quantum source[163], and partially coherent source[164, 165]). Moreover, reinforcing the capacity of information transmission is an advanced pursuit of SPI. Currently, the amplitude, phase, polarization and other characteristics of light have yet to be fully utilized in SPI[166]. By harnessing the potent information processing capabilities of deep learning algorithms and incorporating novel detectors with high time resolution, it will be possible to effectively extract high-dimensional information from compressed echo signals[167-170].

The expectation is that over the next several years, there will be substantial growth in SPI driven by deep learning technologies. The integration of SPI and deep learning not only facilitates the practical application of SPI but also promotes innovation in deep learning architectures. These emerging technologies are poised to revolutionize diverse domains of human activity, ranging from medicine to industry.

**Acknowledgements**

The authors gratefully acknowledge support from the National Key Research and Development Program of China (Grant No. 2022YFA1404201), Natural Science Foundation of China (Grant Nos. 62127817, 6191101445, and 62305239), and the Fundamental Research Program of Shanxi Province (grant Nos. 202203021222133, 202203021222113, 202203021222107, 202203021222104).

**Conflict of Interest**

The authors declare no conflict of interest.